\documentclass[12pt]{article}

\usepackage{epsfig}
\usepackage{cite}
\usepackage{amsmath, amssymb, amsfonts}
\usepackage{color}
\usepackage{latexsym}
\usepackage{graphicx}
\usepackage[colorlinks,bookmarks]{hyperref}
\hypersetup{pdfpagemode=UseNone, pdfstartview=FitH, linkcolor=blue,
           citecolor=red, urlcolor=blue}

\def\be{\begin{equation}}
\def\ee{\end{equation}}
\def\ba{\begin{eqnarray}}
\def\ea{\end{eqnarray}}

\def\bdm{\begin{displaymath}}
\def\edm{\end{displaymath}}
\def\la{~\mbox{\raisebox{-.6ex}{$\stackrel{<}{\sim}$}}~}
\def\ga{~\mbox{\raisebox{-.6ex}{$\stackrel{>}{\sim}$}}~}
\def\bq{\begin{quote}}
\def\eq{\end{quote}}

 at 10truept

\newcommand{\beq}{\begin{equation}}
\newcommand{\eeq}{\end{equation}}
\newcommand{\bea}{\begin{eqnarray}}
\newcommand{\eea}{\end{eqnarray}}
\newcommand{\beqa}{\begin{eqnarray}}
\newcommand{\eeqa}{\end{eqnarray}}

\def\la{~\mbox{\raisebox{-.6ex}{$\stackrel{<}{\sim}$}}~}
\def\ga{~\mbox{\raisebox{-.6ex}{$\stackrel{>}{\sim}$}}~}

\def\ltap{\ \raise.3ex\hbox{$<$\kern-.75em\lower1ex\hbox{$\sim$}}\ }
\def\gtap{\ \raise.3ex\hbox{$>$\kern-.75em\lower1ex\hbox{$\sim$}}\ }
\def\gl{\ \raise.5ex\hbox{$>$}\kern-.8em\lower.5ex\hbox{$<$}\ }
\def\roughly#1{\raise.3ex\hbox{$#1$\kern-.75em\lower1ex\hbox{$\sim$}}}

\begin{document}

\thispagestyle{empty}
\begin{flushright}
December 2020
\end{flushright}
\vspace*{1.2cm}
\begin{center}
{\Large \bf Troubles with Global  Monopoles in Quantum Gravity}

\vspace*{1.3cm} {\large Nemanja Kaloper\footnote{\tt kaloper@physics.ucdavis.edu}}\\
\vspace{.5cm} {\em QMAP, Department of Physics, University of California, Davis, CA 95616, USA}\\

\vspace{1.65cm} ABSTRACT
\end{center}

A theory with a global $O(3)$ symmetry broken at a scale $\eta$ 
admits topological configurations: global  monopoles 
realized by Goldstone fields winding around the
core of false vacuum. One may expect them to behave as heavy, big composite objects, 
difficult to make and therefore mostly harmless. However, after gravity is turned 
on, as long as Equivalence Principle holds, one finds that global  monopoles have a 
negative mass, $M \sim  - \eta/\sqrt{\lambda}$, where $\lambda$ is the field theory coupling. 
This could catalyze an instability in the space of a global monopole, by 
the production of additional global monopole-antimonopole  
pairs along with normal particles, leading to energy production {\it ab nihilo}, if the pair energy is dominated 
by their negative rest masses. 
In a theory with unbroken local Poincar\'e symmetry,
this could lead to a divergent `decay rate' of the global monopole configurations.

\vfill \setcounter{page}{0} \setcounter{footnote}{0}
\newpage

\section{Introduction}

It has been argued for a long time that quantum gravity does not permit 
exact global symmetries (see, e.g. \cite{abbottwise,bds,kmr,holman,klls,polc,bs,hebjmr}). An argument often heard 
is that since black holes do not conserve global charges, and they count among the Hilbert space 
states in quantum gravity, in processes where they appear as virtual states the global charges
cannot be conserved. Instead these processes generate global symmetry violating operators in effective theory.
This argument is appealing, but it seems very difficult to implement it directly, for example, as 
a diagrammatic calculation of a global symmetry breaking operator generated by a loop including 
a virtual black hole. 

More recently, different arguments emerged, invoking weak gravity conjecture (WGC) \cite{nima} and
AdS/CFT \cite{harleyquin,quinharley}. These arguments rest on ideas on how the theory 
of gravity may behave in the UV. In particular WGC is  
interesting since it seems to block trying to realize global symmetries as a zero charge limit of gauge symmetries. 
Yet while there is evidence supporting these ideas, we still do not know
if such specific frameworks for UV completions are unique or generic. 
Hence it is interesting to pursue possible IR avenues  
of conflict between global symmetries and gravity, to see if anything can be shaken out from the
mix of the two. 

An example is provided by an interesting exploration of (in)compatibility of non-compact, 
continuous shift symmetry with gravity in \cite{nicolis}. There, the positivity of energy 
(combined with the implicit assumption that the geometry is static) yields a bound on the total field
variation exterior to a source. This is despite the fact that in flat space the field variations 
can be arbitrarily large. Further examples were provided in, for example, \cite{arthur,draper1,draper2}, where
it was found that gravity appears to obstruct large field changes, which indirectly implies that the 
na\"ive shift symmetry of the flat space limit cannot coexist with gravity. 

Here we will pursue a different approach, which will use only very basic IR properties
of gravity, employing very few ideas about the UV. We will consider theories which have 
a spontaneously broken global $O(3)$ symmetry. Theories with such structures seem 
legitimate in flat space without gravity. They have a nontrivial vacuum manifold, with low energy
Goldstone modes described by a nonlinear $\sigma$-model. In addition to local excitations they also 
admit topological global  monopoles, which correspond to nontrivial Goldstone fields winding around the
region of false vacuum in the core. They exist when the vacuum manifold is isomorphic to the sphere
at infinity. In flat space, global  monopoles seem to be nonperturbative structures which are heavy, 
big composite objects, that tend to be mostly harmless \cite{vilshell}. However even this is not without controversy,
since the issue of their stability has already been debated 
vigorously \cite{goldhaber,benrhie,Bennett:1990xy,turk,ggib,perivo,achucarro,watatori,achucarro2}. 
We will briefly review the status of this
debate and outline the features relevant for our argument. We will work with the symmetry breaking
scale $\eta \ll M_{Pl}$, to avoid any additional issues concerning large field variations\footnote{It is in fact 
known that super-Planckian field values induce topological 
inflation \cite{lintopo,vilentopo}, which in light of the results of 
\cite{chovilen} might even be viewed as a form of instability \cite{choguv}.}. 
Further we will work in weak coupling.

Turning gravity on lets global  monopoles spring several surprises. 
First, in the case of isolated global monopoles, their Goldstone fields support 
a solid deficit angle \cite{barvilen}. This appears to regulate the flat space 
divergent energy density distribution outside of a global monopole. 
Second, thanks to this, global  monopoles have a negative gravitational 
mass, $M \sim  - \eta/\sqrt{\lambda}$ \cite{harari}, where $\lambda$ is the field theory coupling. 
Finally, we will invoke a simple argument using 
Equivalence Principle to infer that as a result the inertial mass of global  monopoles is also
negative. Combining this with unbroken local Poincar\'e symmetry, and using continuity 
of dispersion relations, implies that global monopoles should be described in a local theory as
ghosts, since they reside on mass shells which are inverted
relative to normal matter. To a good approximation they seem to be `hard' at low energies. 

This immediately suggests possible catastrophic consequences, just like in the case of theories with 
hard fundamental ghosts \cite{jimc,hollysundrum,empgar,gia1}: if they are at least 
metastable in local theory, and their rest energy is dominated by 
mass terms, global monopole-antimonopole pairs may provide means to 
catalyze an instability of the background spacetime of a global monopole, due to their pair production along 
with normal positive energy particles. In this case, at least, the positive energy theorem
of \cite{schon,witt} is not directly applicable since the background has a deficit angle. As long as the process amplitude is not exactly zero, and the net energy of the pair is negative, 
the `decay rate' of a global monopole would diverge when the theory has unbroken 
Poincar\'e symmetry in the UV. This would be reminiscent to 
how the flat space stable vacuum appears to reject remnants \cite{lenny}.

Hence we may infer that a consistent theory of quantum gravity 
with unbroken local Poincar\'e symmetry in the UV may not support 
stable global monopoles which arise in field theories with 
spontaneously broken global symmetries. If this is true, 
our example, albeit somewhat special, suggests a glimpse into 
what may go wrong with quantum gravity if we try to push global symmetries onto it. 
An alternative, which would prevent this conclusion, is that the pair energy is never negative
despite the fact that their local rest masses are. For this to be the case, a generalization 
of the theorems \cite{schon,witt} to some spaces which are not asymptotically flat could be
true, although there is no proof at this time. 

\section{Global Monopole Solutions and Their Properties}

Global  monopoles are topological solitons of field theories with 
spontaneous symmetry breaking, which contain an $O(3)$ symmetry, with a potential 
 \be
 V(\vec \Phi) = \frac{\lambda}{4} \, \Bigl( \vec \Phi^2 - \eta^2 \Bigr)^2 \, .
 \label{potential}
 \ee
Because any vacuum at infinity spontaneously breaks $O(3)$ to $O(2)$, the 
theory has massless Goldstone bosons. They are 
crucial for the existence of a global  monopole because 
they give long range fields which wind around the global  monopole core, 
providing the topological charge distinguishing a global  monopole from
the vacuum.  
In such theories there are configurations where the field in the target space 
develops a hedgehog configuration around the core of the false vacuum, 
 \be
 \vec \Phi = f(r) \, \frac{\vec r}{r} \, ,
 \label{hdegehog}
 \ee
wrapping around it (in the example above, exactly once) as it interpolates to the degenerate
manifold of true vacua at infinity (for a review see \cite{vilshell}). 
As the Goldstones range out to infinity in a global  monopole 
configuration, interpolating between the different true vacua far away, they yield 
stress energy tensor which vanishes only as \cite{vilshell} 
\be
T^0{}_0 = - T^r{}_r \sim \frac{\eta^2}{r^2} \, ,
\label{tmunu}
\ee
while the angular components vanish much faster. As a result,
individual global  monopoles formally seem to have a divergent field energy, 
\be
E = 4\pi \int^R dr \, r^2 \, T^0{}_0 \sim 4\pi \eta^2 R \, . 
\label{energy} 
\ee
Hence individual global monopoles cannot
be produced in flat space unless infinite energy initial conditions are used. 
Thus one may imagine that due to such huge energy 
in the global  monopole field, its mass, and therefore its inertia, should be very large.
However this single source solution can be regulated by adding a global antimonopole, which is a parity image
of a global monopole, at a distance $\sim R$ from the global monopole \cite{vilshell}. In that case the 
energy (\ref{energy}) will be cut off at infinity, since the antimonopole serves as a sink
of field lines, at a finite distance from the source. This yields a faster dilution of stress energy
far away.

Early on it has been asserted that a global  monopole may be 
unstable \cite{goldhaber} (see also \cite{benrhie,Bennett:1990xy}) 
since it can unwind by itself, thanks to an angular perturbation of the spherical isopotential 
shells around it. A small perturbation of the global monopole on, say, the North Pole 
has a tendency to grow in the axial direction, and pull in the Goldstone field lines towards the 
vertical axis, where they can unwind. 

A more detailed analysis 
has shown that this instability really induces a force acting on the global monopole, 
accelerating it in the direction of the perturbation \cite{benrhie,turk,ggib,perivo}. We could imagine 
this configuration of a ``leashed" global monopole as a global 
monopole-antimonopole  pair separated by 
a large distance. The array breaks spherical symmetry, since the 
presence of a global antimonopole introduces a direction, and so the field lines tend to align 
with it and form a global `flux tube' \cite{turk,ggib}. The initial realignment starts slowly due to a large separation, 
but then the constant force pulls the two together \cite{perivo}.  

Thus we assert that 
the instability observed in \cite{goldhaber} can be viewed as a signal of the onset 
of the process of global monopole-antimonopole annihilation in the large initial distance 
limit \cite{benrhie,turk,perivo}. Moreover,
although the topological charge is accounted for by the boundary conditions at infinity, 
the energy barrier between configurations with a different charge seems to be finite \cite{achucarro}. 
This suggests that the number of global monopoles could
be `rearranged' with a finite energy investment \cite{achucarro} 
despite the apparent divergence of field energy (\ref{energy}).

More careful examinations with gravity turned on reveal interesting surprises. 
To see how the energy (\ref{energy}) behaves, we may use General Relativity as a probe. 
It turns out that the leading 
order sources in $T^\mu{}_\nu$ do not yield any local gravitational forces. Instead, the leading order $T^\mu{}_\nu$ 
which had yielded the divergence (\ref{energy}) in flat space now sources a global solid 
deficit angle of the space exterior to a global  monopole \cite{barvilen}. 
The metric of a global  monopole outside of the core is 
\be
ds^2 = - \Bigl(1- 8\pi G_N \eta^2 - \frac{ 2G_N M}{r}\Bigr) \, dt^2 
+ \Bigl(1- 8\pi G_N \eta^2 -\frac{ 2G_N M}{r}\Bigr)^{-1}  dr^2 
+ r^2 d\Omega_2 \, .
\label{barvimet}
\ee
The solid angle surrounding a global monopole at infinity is $\Delta \phi = 4\pi (1- 8\pi G_N \eta^2)$, 
which follows from
the leading order behavior of the metric as $r \rightarrow \infty$. Thus the leading order
energy is effectively degravitated.
Locally, all it does is enhance the value of Newton's constant in the space outside of a global  monopole. 
This is obvious from Gauss' law at infinity, where the angular deficit implies that there is stronger 
fields for a fixed source mass since there is  less  spatial volume for the field lines to dilute \cite{kalkil}. 
The long range effects of $T^\mu{}_\nu$ persist only as the deficit angle \cite{barvilen,harari}. 
They are important for the understanding of a global  monopole-antimonopole  pair creation
as we will argue later; they are crucial for understanding their local gravitational and inertial 
properties. 

At the subleading level, gravity reveals an even greater surprise \cite{harari}: a global  monopole has a 
{\it negative} gravitational mass! Technically, this rather odd feature can be 
understood as follows. Inside a global  monopole
core the field configuration is near the false vacuum $\Phi = 0$, where the energy 
density is $\rho \sim V \sim \lambda \eta^4/4$, which behaves like a positive 
cosmological constant, exhibiting repulsion of normal matter due to the 
negative pressure $p \sim - \rho$, which induce forces $\propto (\rho+3p) < 0$.
The interior geometry is approximately 
given by de Sitter static patch metric,
\be
ds^2 = - \Bigl(1- H_{core}^2 r^2 \Bigr) \, dt^2 +  \Bigl(1- H_{core}^2 r^2 \Bigr)^{-1} dr^2 + r^2 d\Omega_2 \, ,
\label{dsmetr}
\ee
where
\be
H_{core}^2 \sim \frac{8\pi G_N V}{3 } \sim \frac{2\pi G_N \lambda \eta^4}{3} \, .
\label{dscurv} 
\ee
Because the exterior metric has a deficit angle, the interior and exterior can be 
smoothly matched at the ``surface" of the global  monopole, where the metrics and their 
derivatives continuously connect. This yields \cite{harari}, up to corrections $\propto 8\pi G_N \eta^2 \ll 1$,
\be
r_0 \simeq \frac{2}{\sqrt{\lambda} \eta} \, , ~~~~~~~~~~~~~~ M \simeq - \frac{16\pi \eta}{3\sqrt{\lambda}} \, .
\label{sizemass} 
\ee
The mass in the exterior metric must be negative by Birkhoff's theorem, 
since the interior mass \cite{harari,schuck} is
$M_{core} \sim (\rho_{core} + 3 p_{core}) \,  {\rm Vol}_{core}  \sim - 2 \rho_{core} \, {\rm Vol}_{core}$. 
Importantly, also note that as long as
\be
\frac{2G_N |M|}{r_0} \sim \frac23 8\pi G_N \eta^2 \la 1  \, ,
\label{horizons}
\ee
or so as long as $8 \pi G_N \eta^2 \la 1$, the description of a monopole using 
flat space physics should be reliable, 
since gravity is too weak in its core. Clearly, we are in this limit when $8\pi G_N \eta^2 \ll 1$. 
Further note that by combining Eqs. (\ref{sizemass}), 
\be
r_0 |M| = \frac{r_0}{\lambda_{compton}} \simeq \frac{32\pi}{3\lambda} \gg 1 \, ,
\label{compton}
\ee
which shows that in perturbation theory, $ 8 \pi G_N \eta^2 \la 1$, $\lambda \ll 1$, the 
global monopole core size is much larger than its 
Compton wavelength. So, global monopoles are classical, extended 
objects\footnote{As a technical aside, to fully describe them we need to keep the radial mode 
in the theory. Since the mass of the radial mode of 
$\vec \Phi$ is $m_\Phi \sim \sqrt{\lambda} \eta$, we find $r_0 m_\Phi = {\cal O}(1)$,
implying that global monopoles are about the same size as the Compton 
wavelength of the radial scalar modes, integrated out in the true vacuum. 
Hence global monopoles explore the physics above the cutoff of the 
perturbative $\sigma$-model describing only the Goldstones, since the cutoff
of the low energy effective theory is set by $m_\Phi$.}, which as long as 
$ 8 \pi G_N \eta^2 \la 1$, $\lambda \ll 1$ have gravitational properties that 
can be consistently described using linearized General Relativity. 

But this is not all: 
as long as gravity is governed by General Relativity, which obeys Equivalence Principle, 
the global  monopole inertial mass $M_{inertial}$ must also be negative since it must be equal 
to its gravitational mass $M$. To see this, imagine 
a probe particle near a global  monopole, in free fall in the geometry with the metric (\ref{barvimet}), following
the geodesic equation 
\be
\ddot x^\mu + \Gamma^\mu_{\nu\lambda} \dot x^\nu \dot x^\lambda = 0 \,  .
\ee
Let the particle stand to the right of the monopole, moving slowly and being separated a distance $r$. 
Because the global  monopole gravitational mass $M$ is negative, the particle experiences a
repulsive acceleration controlled by the global  monopole's negative gravitational mass, 
as the particle mass cancels 
out of the geodesic equation. So the force pushing the particle away 
from the global  monopole is $\vec F \sim - mM \vec r_0/r^2 = m |M| \vec r_0/r^2$. 

Now: change coordinates from the rest frame of the global  monopole to the rest 
frame of the particle, and look at the motion of the global  monopole 
governed by the gravitational field of the particle, with a Schwarzschild metric 
of the mass $m$. To the leading order,  the global  monopole follows geodesic 
equation in the new coordinate system, experiencing an attractive acceleration 
induced by the positive particle mass -- since this time the global monopole's inertial mass 
cancels from the geodesic equation, the gravitational field and the acceleration imparted by it is the
same for all probes, controlled only by the particle's positive mass $m$, as follows from 
Equivalence principle. 
The force pulling the global  monopole towards the particle (on the right of the monopole) is 
$\hat F \sim m M_{inertial} \,  \vec r_0/r^2$. Again by Equivalence 
Principle -- since the motion of the particle in the global  monopole field must be 
the same as the motion of the global  monopole in the particle's field -- the 
forces $F$ and $\hat F$ must be equal and opposite \cite{einstein,gibrub}, and 
hence $M_{inertial} = - |M| = M \equiv M_{gravitational}$, as claimed. Therefore 
\be
M_{inertial} \equiv M \simeq - \frac{16 \pi \eta}{3 \sqrt{\lambda}} \, .
\label{inertialmass} 
\ee

The calculation of the mass $M$ can be made more precise by generalizing the notion
of the ADM mass in the presence of the deficit angle, which 
changes the boundary conditions at infinity. 
The generalization of the ADM mass for spacetimes with solid deficit angle \cite{nucamendi,nuc} confirms  
the intuitive results after the renormalization of $G_N$ by the deficit angle. 
This leaves us precisely with the result of (\ref{sizemass}) to the leading order in the 
$8\pi G_N \eta^2$ expansion \cite{harari,nucamendi,nuc}, allowing the mass of a 
global monopole to be negative. 

The price, of course, is the deficit angle, amounting to the 
alteration of the boundary conditions at infinity. In flat space, 
the net global monopole number -- ie the total winding number -- enumerates superselection 
sectors. To create a single global monopole would presumably be obstructed by the need to `bring' the charge from 
infinity, and so spend an infinite amount of energy, since Goldstone fields dilute too slowly far away.
However, these superselection sectors seem to only be separated by finite energy barriers \cite{achucarro}.  
This, at least, suggests that a global monopole-antimonopole pair can be formed readily. 
In flat space without gravity, if the initial state is the Minkowski vacuum, or even a configuration 
with a global monopole on shell, a global monopole-antimonopole  pair could  
be created quantum-mechanically if the conservation laws permit it. 
In this limit all the energies are dominated by positive rest masses, by the 
theorems \cite{schon,witt} in asymptotically flat spaces. So the pair will quickly fall in 
and annihilate -- with a zero net energy release -- existing a mere moment only because 
of the uncertainty principle. If so, the flat space vacuum would be stable, since these events would merely 
be one of the many quantum fluctuations that come and go. The diagrams representing their contributions to any
observables would be disconnected bubbles, which factor out. In this argument the 
gravitational effects are completely ignored.

Turning gravity on brings out the negative global monopole masses and deficit angles. 
Having objects with negative masses, with all local energy densities being positive, is a very unsettling proposition. 
Already \cite{harari} note that if these configurations really existed, they could 
lead to the formation of Bondi dipoles \cite{bondi}. 
After noting this option, \cite{harari} step back and 
dismiss it on the grounds that global  monopoles are topological objects
which are not `isolated point particles, but extended sources.' 
A single global monopole is indeed an extended object in the strict geometric sense: it affects the boundary 
conditions at infinity. Yet at high energies a global monopole can unwind as 
the symmetry is restored. The charge and the field energy would then flow away to infinity. 

On the other hand, global monopole-antimonopole  pairs extend far less, with their long distance
effects diluting away more quickly. This naturally regulates the energy divergences
far away. At distances much larger than the global monopole-antimonopole separation, the net charge of the
pair will look to be zero, which means that the field strengths of the Goldstones must diminish faster
than in the case of a single global monopole. Therefore in a quantum mechanical
process global monopoles could be created as pairs and so a net global monopole 
number should not even change\footnote{Such events could be extremely rare, but if 
the rates are not zero the problems will arise.}. However the numbers of global monopoles and
global antimonopoles may fluctuate. 

This should survive the onset of 
gravity; in pair creation, the curvature deformation mimicking 
deficit angle may be induced locally near individual sources after the
pair emerges, but will not extend to infinity as long as the field energy 
between the pair is finite. This local distortion of the geometry
may subsequently propagate farther -- or not -- depending
on the fate of the pair. However in practice, in the linearized gravity regime applicable to
our analysis due to choosing $8\pi G_N \eta^2 \ll 1$, $\lambda \ll 1$, inside the region of space 
where the pair forms, and close to the global monopoles, they will behave as
negative mass objects when probed by normal matter for all 
practical intents and purposes. 

We contend that this may allow the global monopole-antimonopole  pair creation 
to drive an instability of a configuration {\it initially containing a global 
monopole on shell} through particle production. 
As we stressed above, the formation of a global monopole-antimonopole  pair out of the flat space 
vacuum is not prohibited a priori if all conservation laws are satisfied. Since
Goldstone fields are cut off, this is a finite energy configuration. As we 
stated above, this configuration should be interpreted 
as a pair of global monopoles connected with a `flux tube'. This is motivated by the 
studies of Goldhaber's instability \cite{goldhaber,benrhie}, whereby a global monopole
unwinds directionally before annihilating with a global antimonopole \cite{benrhie,turk,ggib,perivo}. Since
the global antimonopole is needed to regulate the Goldstone field energy 
(\ref{energy}) anyway, it is natural to interpret Goldhaber's 
instability as the tendency of Goldstone fields to collapse to a flux tube \cite{turk,ggib}. We maximize its energy
by assuming no energy loss in the field rearrangement, so that the interaction energy mediated by the 
flux tube is given by the regulated expression (\ref{energy}). This way we reproduce the formula
for the constant force acting on a global monopole (as already noted in \cite{ggib}). 

We can then view 
the global monopoles at the end points of the tube as locally 
negative mass objects, when we get near them (see also \cite{gia2} for a similar point of view). 
The question then is, what is the energy of the pair, and could it be negative. 
This question is subtle\footnote{We thank A. Vilenkin for useful communications about this issue.}. 
Although it may appear that we could make the energy negative for a pair even in vacuum, when they 
are initially sufficiently close, so the flux tube between them is short, and they 
are moving, so their kinetic energy may add to their negative 
rest masses, the overall global monopole-antimonopole resonance energy in asymptotically flat space
is prohibited from being negative by the theorems of \cite{schon,witt}. This would block the instability in the vacuum.

However if the background already contains global monopoles, and 
therefore has a deficit angle, the case is not so clear cut. The theorems \cite{schon,witt} do not
automatically guarantee that the pair energy must be positive. 
The total energy of the
system might not be positive definite due to different boundary conditions at infinity. 
In this case, if a global monopole-antimonopole pair pops out in 
a background already containing a global monopole, 
with an overall negative energy, it must be accompanied by at least one positive 
mass particle-antiparticle pair, such that the total system satisfies 
energy-momentum conservation enforced by local Poincar\'e invariance -- that the net energy change is zero. 
Hence the production rate may be kinematically allowed. Once the positive mass particles get on shell,
they will fly away, since they can be light -- or even massless, being photons or even gravitons themselves. 

In many cases the global monopole-antimonopole pairs which are created will fall in and unwind away, absorbing
some of the positive mass particles floating aroud; in some, 
they may even end up wiggling around for a while and increasing their (negative) kinetic energy by emitting 
positive mass particles. But their ultimate fate past the fact that they could be created seems largely irrelevant 
for the instability. If they are on shell, the positive mass particles {\it will} escape regardless of 
the fate of the global monopole-antimonopole  pair, and stress-energy may 
flow out of the space which is asymptotically 
locally the same as the vacuum, except for the deficit angle. 
When local Poincar\'e symmetry is unbroken in the UV, the phase space for the outgoing decay products 
has divergent volume just like similar processes for the production of hard elementary 
ghosts, where the energy-momentum conservation holds  \cite{jimc,hollysundrum,empgar}. 
As a result, as long as the decay rate is not exactly zero, it should diverge. 
The suppression due to the compositeness 
of the global monopoles may make their pair production rate per 
unit phase space volume exponentially small, but as long as it is not
zero the disaster is lurking. All it takes is a pair with a net negative energy and local Poincar\'e invariance
in the UV.

An alternative outcome is that even though the space has a deficit angle, and so is not asymptotically flat,
the results of \cite{schon,witt} still hold and exclude negative energy pairs. If this is true then our discussion
below may be viewed as an explanation why \cite{schon,witt} need to extend to this case, despite the
global redefinition of the ADM mass \cite{nucamendi,nuc}. In this case reconciling the negative mass of 
an isolated global monopole with the positive energy of a pair would be quite interesting to see.

\section{The Global Monopoles' Fizz}

Now we turn to the details of how global  monopoles could trigger the instability in the space surrounding them. 
As we noted in the beginning, we will be working with safely subcritical global monopoles, 
where $8\pi G_N \eta^2 \ll 1$, $\lambda \ll 1$ so that we can for the most part ignore the 
deficit angle. Locally, it only renormalizes $G_N$ in this limit; however it is crucial for
the existence of negative masses. 

Consider now a global monopole-antimonopole pair, separated by a distance $R$, in a locally asymptotically 
flat space of a distant global monopole. 
The global monopole far away sets up the deficit angle at infinity. 
In the least favorable case to our argument, when the force is attractive, since it 
is constant \cite{vilshell,perivo}, the time it takes for the pair to annihilate is controlled by the initial separation,
\be
\Delta T \sim \sqrt{\frac{R |M |}{4\pi  \eta^2}} \, .
\label{lifetime}
\ee
At distances $r > r_0$ we will treat individual global monopoles comprising the pair as hard particles, 
with negative masses $M$ of Eq. (\ref{inertialmass}). 
To the leading order, when they are close enough their total energy 
may be dominated by the kinetic terms due to negative masses. 
The negativity of the mass combined with unbroken 
local Poincar\'e symmetry implies that global monopoles 
would behave as {\it ghosts}, with negative energy 
\be
E = - \sqrt{\vec p^2 + M^2} \, , 
\label{energyrel}
\ee
for a given momentum $\vec p$. To see this, imagine
an observer boosted relative to a global monopole at rest. In the rest frame of the
moving observer, it appears that the global monopole 
is moving in the opposite direction, with the relativistic energy $E \ne M$. By continuity of boosts,  
we can dial down the relative velocity $\vec v$ 
to zero smoothly, which means that in the limit $\vec v = 0$ we must reproduce 
$E = -M$. Therefore $E = - \sqrt{\vec p^2 + M^2}$. This means that the mass shells of the global 
monopoles are inverted relative to normal particles.

Next, we estimate how far the monopoles can be so that the net energy 
is negative, using the energy balance 
formula based on (\ref{energy}). Note that if the initial state was empty Minkowski space
the net energy would have to be non-negative by \cite{schon,witt}. However if there is a deficit angle due to a 
global monopole far away, this need not be true. Note also that the formula (\ref{energy}) is
at best only an approximation in a conical space, but we will nevertheless use it to illustrate the issues. 
So for a global monopole-antimonopole at rest, with negative masses (\ref{sizemass}), to the leading order
we take the excess energy over the energy of the background monopole far away to be roughy given by
the flat space formula for the flux tube energy, but with negative monopole masses,
\be
E_{total} \simeq 4\pi \eta^2 R + 2M = 4\pi \eta^2 R - \frac{32 \pi \eta}{3 \sqrt{\lambda}} \, ,
\ee
and allowing for it to be negative if the monopole-antimonopole separation $R$ is small enough,
\be
R_{cr} \la \frac{8}{3 \sqrt{\lambda} \eta}  \simeq \frac43 r_0 \, .
\label{rcr}
\ee
The distance $R_{cr}$ is just-so larger than the global monopole core size, 
meaning the global monopole-antimonopole pair are in contact. They will 
not unwind right away, but may do so, at a rate depending on $r_0$. 
Still, even this may suffice to trigger a cataclysm. Note that as we take $\lambda \rightarrow 0$
the danger from long lived negative energy pairs seems to grow. 

Moreover, the monopoles are initially moving, having additional kinetic energy 
so that $-M \rightarrow -M/\sqrt{1-v^2}$, so the critical separation might be larger, 
and the net energy of the pair might still be negative. 
This opens a window for a global monopole-antimonople separation to be at least a few times 
larger than $r_0$, improving the
approximation where the monopoles are treated as hard ghosts. 
Thus when the pair are moving initially, by causality 
they will survive for at least a time 
$\Delta T \sim ({R_{cr}}/{\sqrt{\lambda} \eta})^{1/2} >  {\rm few} \times r_0$.
Such global monopole-antimonopole pairs would be approximately on shell: 
quantum production of photons will be faster than global monopole-antimonopole annihilation. 
Hence these global monopole-antimonopole pairs ought to be included among those states which the 
space that contains a global monopole can decay into, at least as metastable resonances.

Since in locally Poincar\'e invariant theories we must enforce energy conservation,
additional positive energy particles would be produced, 
such that the total energy change is zero \cite{jimc,hollysundrum}.
A simple cartoon for a process describing this is a spontaneous creation of a bound system of 
scalar quanta forming the global monopole-antimonople pair along with a swarm of free 
streaming `photons' mediated by gravitational interactions (or any other particle-antiparticle pair, 
due to the universality of gravity, including gravitons or even the $O(3)$ Goldstones). 
The global monopole setting the background can be taken to be far enough that its only 
effect on local physics is via the 
deficit angle, which invalidates positive energy condition. Otherwise the dynamics is as in any locally
flat space. Treating the background as a `vacuum', the 
total process would involve many different internal scalar excitations, whose 
number would be at least as large as $M/m_\Phi \sim M r_0 \gg 1$, with an 
arbitrary number of external photons popping out, as long as the net energy is zero. 
A channel with a given number of scalars and two photons is sketched in Fig. 1. 
\begin{figure*}[thb]
\centering
\includegraphics[scale=.55]{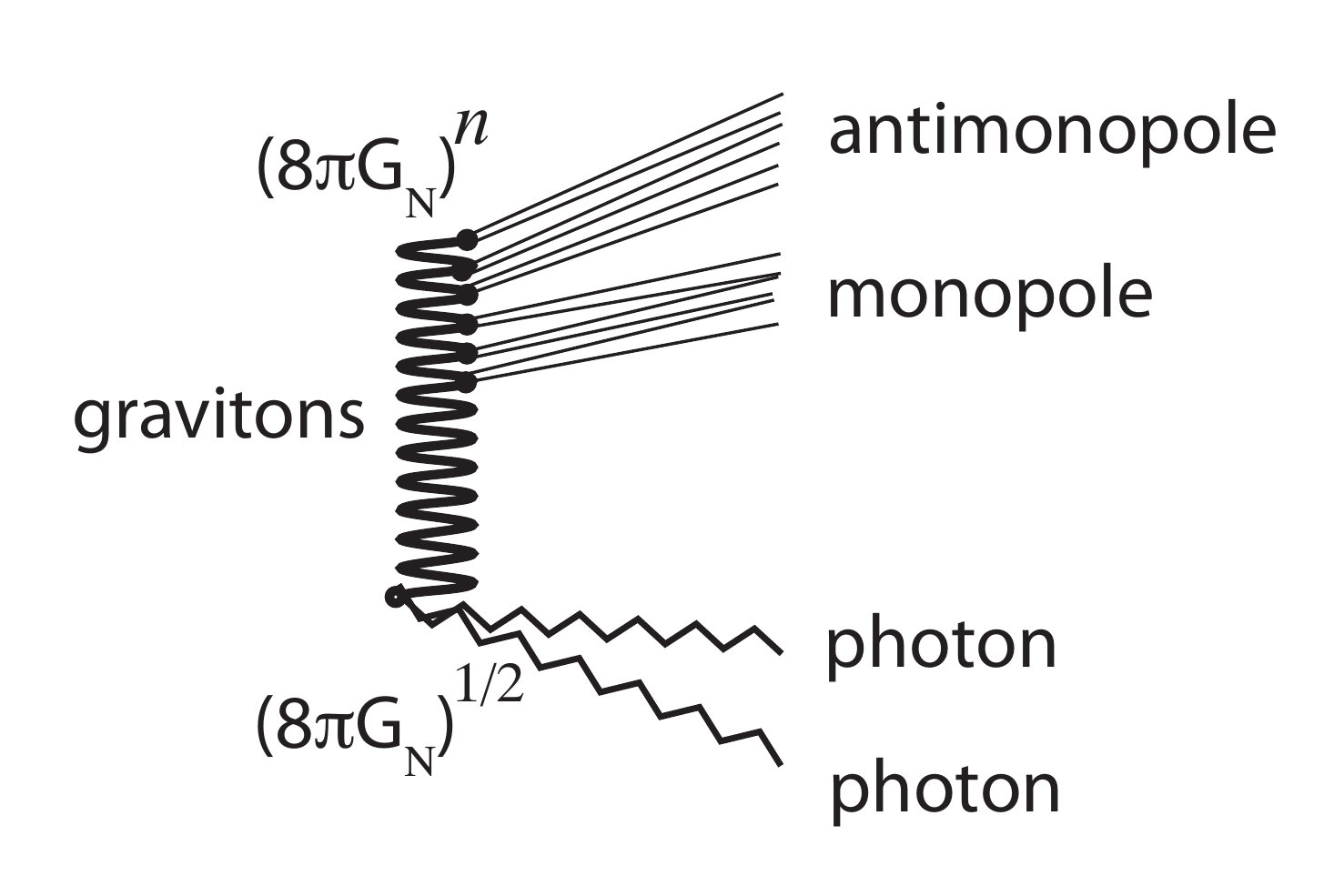}
\caption{One of many channels for producing a global monopole-antimonopole  pair + two photons, 
mediated by gravity, far from an initial global monopole on shell. Since global monopoles are composite, at least 
$\sim 2n \sim 2 Mr_0$ scalar `quanta' are required to make up the pair. }
\label{fig1}
\end{figure*}

Clearly, there is an exponentially large number of such channels. 
The total amplitude of the process is obtained by summing amplitudes for each channel.
A large number of channels means that they would sum up 
mostly out of phase, yielding destructive interference and suppressing the amplitude. 
The rate would be effectively suppressed by compositeness of the global monopole-antimonopole  pair. 
However even an exponential suppression $\sim \exp(-S)$ 
is not enough to stop the disaster, as long as the pair energy could be negative. 

Indeed, this could trigger a catastrophe because 
the Lorentz-invariant phase space for such processes is infinite. This 
is regardless of whether global monopole-antimonopole  pair 
end up going to infinity, or not. 
The important point is, there is no stopping the photons, 
which would carry energy away with them. If realized, this leads to 
asymptotic space instability, resulting in energy production {\it ab nihilo}. 

Although the initial states which contain a global 
monopole and  deficit angle break the local Poincar\'e spontaneously, 
this breaking is in the IR and does not help with UV divergences. 
The reason that the phase space diverges 
for a process like
$$
| 0_{asymptotic} \rangle \rightarrow |M \bar M \, \gamma \gamma \ldots \, \rangle
$$
is that the global monopole masses are negative, due to deficit angle. 
The divergence can be readily inferred from the transition amplitude following the analyses in 
\cite{jimc,hollysundrum,empgar,gia1}. Let's consider the simplest divergent contribution coming from 
the process $0_{asymptotic} \rightarrow M \bar M \, \gamma \gamma$ whose 
one channel is depicted in Fig. \ref{fig1}. 
The contribution to the decay rate
of the asymptotically flat space far from the global monopole source 
via the production of a global monopole-antimonopole  pair plus two photons is 
\ba
\Gamma &\ni & \int d^4 p_1 d^4 p_2 d^4 k_1  d^4 k_2 \, | A |^2 \, 
\delta(p_1^2 - M^2) \delta(p_2^2 - M^2) \delta(k_1^2) 
\delta(k_2^2) \times ~~~~  \nonumber \\ 
&&~~ \times  \, \theta(-p_{10}) \theta(-p_{20}) \theta(k_{10})
\theta(k_{20}) \delta^4(p_1 + p_2 + k_1 + k_2) \, , 
\label{gammadef}
\ea
Here $A$ is the transition amplitude for the process 
$0 \rightarrow M \bar M \, \gamma \gamma$, defined in the standard way by 
$\langle p_1,p_2,k_1,k_2|0_{asymptotic}\rangle={ A}(p_1,p_2,k_1,k_2)\
\delta^{(4)}(p_1+p_2+k_1+k_2)$ (thanks to the local Poincar\'e symmetry),
$p_j$ are the global monopole $4$-momenta and $k_j$ are the photon $4$-momenta,
and the signs in the step functions $\theta(-p_{j0}), \theta(k_{j0})$ 
take care of the mutual reversal of the mass shells,
accounting for the negative inertial mass (\ref{inertialmass}). 

To proceed, it is convenient \cite{empgar} to rewrite the integral in (\ref{gammadef}) by introducing an auxiliary 
$4$-momentum $P$, and integrate over it to enforce $P = p_1+p_2 = -k_1-k_2$, which executes  
$4$-momentum conservation in two steps. 
This yields \cite{empgar}
\be 
\Gamma \ni \int d^4 P \, \gamma(P) \, , 
\label{rate1}
\ee
where 
\ba
\gamma(P) &=& \int d^4 p_1 d^4 p_2 d^4 k_1  d^4 k_2 \, | A |^2 \, 
\delta(p_1^2 - M^2) \delta(p_2^2 - M^2) \delta(k_1^2) 
\delta(k_2^2) \times ~~~~  \nonumber \\ 
&&~~ \times  \, \theta(-p_{10}) \theta(-p_{20}) \theta(k_{10})
\theta(k_{20}) \delta^{(4)}(P+k_1+k_2)\
\delta^{(4)}(P-p_1-p_2) \, .
\label{rate2}
\ea
In this integral, the auxiliary variable $P$ appears as an external momentum. The local Poincar\'e symmetry
then asserts that -- since there are no preferred directions in the UV -- $\gamma$ can depend only on the magnitude 
$P^\mu P_\mu = -s$. Note that this is a consequence of the {\it Lagrangian} of the theory. 
Then changing the 
integration measure in (\ref{rate1}) to $s$ and $\vec v = \vec P/\sqrt{s}$ yields
\be
\Gamma \ni  \int \frac12 \frac{d^3 \vec v}{\sqrt{1+\vec v^2}} \int ds\ s\
\gamma(s) \, .
\label{rate3}
\ee
This change of variables is analogous to spherical polar coordinates in $4D$. However 
the difference is that $P$ is a vector in a $4D$ Minkowski space, and thus in the decomposition
$P^\mu P_\mu = P^2(1-\vec v^2) = - s$ the manifold of $\vec v$ is not compact. Indeed,
using $d^4 P = dP P^3 dH_3 = ds s \, dH_3/2$, and realizing that $dH_3$ is 
the measure on the mass shells defined by $v_0^2 = 1 + \vec v^2$, we see that 
$\int dH_3/2 = \frac12 \int dv_0 \, d^3 \vec v \, \delta(v_0^2 - 1 - \vec v^2) 
= \int \frac12 \frac{d^3 \vec v}{\sqrt{1+\vec v^2}} \sim 2\pi \int d v \, v$.  
If this were an integration over euclidean vectors, it would 
correspond to integrating over a hemisphere from the pole to 
the equator, which would give a finite answer since the sphere 
is compact\footnote{In that case, the sign under the 
square root in the denominator of (\ref{rate3}) would be negative.}. 
However, the Lorentzian signature implies that the integration is 
over an infinite range of the lengths of $\vec v$.
Thus as long as $|A|^2 \ne 0$ the integral diverges. 

We can understand this by considering
$4$-momentum conservation $p_1 + p_2 + k_1 + k_2 = 0$. Since the photons
are null, $k_1^2 = k_2^2 = 0$. For the global monopoles, we have $p_1^2 = p_2^2 = M^2$. 
Rewriting the conservation equation as the expression for $p_1$ in terms of the rest then
yields, after a simple manipulation, $p_2 \cdot k_1 + p_2 \cdot k_2 +  k_1 \cdot k_2 = 0$. 
Next, we can perform a boost to the rest frame of $p_2$, where $p_2 = (M,0)$. In this frame
$k_j \cdot p_2 = M E_j$. Using the projection constraint and defining the angle $\theta$
between the two photons via $\vec k_1 \cdot \vec k_2 = E_1 E_2 \cos(\theta)$, we finally obtain the
equation
\be
\cos(\theta) = 1+ M(\frac{1}{E_1} + \frac{1}{E_2}) \, .
\label{cos}
\ee 
Since the photons are on the positive mass shell, the term in the parenthesis is positive definite. Thus
if $M > 0$, there would be no solutions for $\theta$, since the RHS would exceed unity. However
since the global monopole mass is negative, $M < 0$, the RHS 
will be smaller than unity at energies $\ga {\rm few} \times |M|$,
and the kinematical constraint (\ref{cos}) will be satisfied for some scattering angle $\theta$. 
This will happen not only in the rest frame of $p_2$, but in any frame boosted relative to it, 
with the same value of $s = - (p_1 + p_2)^2$, which must also be
included among the final decay states. Since there is infinitely many such frames the rate will diverge. This last
step is precisely the integration over $\vec v$ above. 

As long as we maintain that the local Poincar\'e symmetry is unbroken, 
the only way to reverse this conclusion seems to be to have $|A|^2 \equiv 0$, or to block  
the global monopole-antimonopole pair from negative values. 
But this is precisely the point we want to make: 
as long as a global  monopole-antimonopole pair with negative 
energy can be formed in the background with an initial
global monopole,  
however unlikely this outcome might be, the decay rate will blow up due to unbroken local Poincar\'e invariance. 
It appears that one way to accomplish this switch-off
of $A$ is to decouple gravity completely, sending $M_{Pl} \rightarrow \infty$. The world 
without gravity is not what we are after, however. 

We would be remiss to ignore the possibility that the positive energy theorems from asymptotically flat space
\cite{schon,witt} somehow extend to the conical space of a global monopole. This could preclude 
the situation that the global monopole rest masses ever dominate the flux tube energy. We
cannot exclude this, although it is difficult to see how this could occur\footnote{Note that if we could take the 
limit $\lambda \rightarrow 0$, we might think the negative mass monopoles would decouple (they would in flat space without 
gravity). However this 
would enhance the global symmetry from $O(3)$ to the full symmetry group of the $3D$ Euclidean space 
by adding three 
commuting continuous translation shift symmetries. From the point of view of quantum gravity this would seem 
to be an even worse outcome.} for all values of $\lambda$. Further if this is the case, then reconciling such a statement
with the negative mass of 
an isolated global monopole which follows from the redefinition of the ADM mass to conical spaces \cite{nucamendi,nuc}
would be quite interesting to see.

Alternatively, if the theory has a UV completion which is not locally Poincar\'e invariant, the 
instability rate can be cut off and global monopoles may coexist with gravity. An example
might be global monopole-like configurations in `emergent gravity' models involving 
superfluid Helium \cite{volovik}. Since the UV completion is just the standard nonrelativistic 
quantum mechanics, the absence of local Poincar\'e invariance in the UV introduces a cutoff
to the integrals contributing to the decay rate. 

Another alternative is to declare that stable global monopole states cannot exist in a world with gravity. 
E.g. those states must include an arbitrary number of excitations and hence may not be normalizable 
in the standard way. This would be in line with the lore that global symmetries
cannot coexist with quantum gravity. Thus our example, while special to an extent since it requires a 
global spontaneously broken symmetry, may be a piece of evidence supporting
this lore. 

Simply put, in the presence of a spontaneously broken 
global symmetry in a world with General Relativity the solitonic structures supported by spontaneous breaking 
of global symmetry provide long range Goldstone fields sourced by 
the topological charge. They backreact on the geometry and induce a local distortion around 
the global monopole cores, which 
behaves like a `cavity' that traps a section of 
de Sitter space in the core of the monopole. This region is nonsingular, smooth and completely `naked' 
to the exterior: it is not hiding behind a horizon, and 
is not covered by a shell of material. Thus the relativistic pressures of the vacuum
energy, which also gravitate, allow the mass 
of the configuration to be negative. A deficit angle at infinity may permit configurations of 
multiple such objects where the negative masses dominate. If so, without a strict cutoff, the disaster looms. 
A space of a single global monopole becomes crowded quickly.

\section{Summary} 

In summary, we have argued that global monopole configurations in General Relativity with unbroken
local Poincar\'e invariance might be unstable to quantum effects. They have a deficit angle and 
as a consequence, a negative 
mass, both gravitational and inertial \cite{harari,einstein}. 
Although their production rate may be very small,
if they can be on shell when paired up with global antimonopoles, and have the pair energy which is 
negative, which might happen when an initial global monopole is 
already present, obstructing the theorems of \cite{schon,witt}, 
normal particles which conserve total 
$4$-momentum could be emitted. The rate for these processes is divergent
because the unbroken local Poincar\'e symmetry in the UV implies that the 
phase space accessible to the out states is infinite.

One possible `obstruction' to our argument is that the result of \cite{schon,witt} might still be
valid despite the fact that a global monopole sets up a conical space. If so, our discussion could be taken 
as an indirect argument to this extent, since otherwise a rapid instability will occur. A more general positive energy theorem would exclude
the pair production instability. If on the other hand even a single pair might pop up with a negative energy,
this would raise an obstacle to having global symmetries in a 
QFT coupled to universal gravity, with 
the only invocation of UV being to require that the single-excitation states in asymptotically locally flat space 
exist forever\footnote{A pragmatic demand might only be that the 
locally flat space, and isolated individual monopoles, exist for $\ga 10^{10}$ years, 
to fit the observations. If so, the finiteness of the observable universe
might help regulate divergence of the decay rate.} 
and that the theory is locally Poincar\'e invariant. Everything else 
rests on the IR properties of gravity.

If the latter option is the outcome, then how does gravity exorcise global monopoles, and a global symmetry needed to produce them, 
out of the spectrum of the theory? At this point, this is a wide open question. 
One option is that those states simply do not belong to the theory since they would not 
be normalizable in the usual sense, as the number of particles diverges. 
Yet another possibility is that gravity further destabilizes global monopole-antimonopole pairs (unlike individual 
global monopoles which seem to be more stable
in classical gravity than without it \cite{watatori}), and hence they never get on shell with 
negative energy. Either way, the global
monopoles merely flag disconnected superselection sectors determined by boundary conditions
at infinity. This is 
unlike what happens with topological couplings in gauge theories, where gauge symmetry allows sectors
with fixed winding numbers to mix, since after all there is no gauge symmetry here. Restricting
the theory to a single sector would effectively break global symmetry. 

A question then is how this could manifest 
locally. Perhaps quantum gravity, via, e.g. wormhole
effects, such as those proposed long ago in  \cite{abbottwise,kmr,holman,klls}, induces explicit global 
symmetry breaking terms such as $\sum_j (\vec A^j \cdot \vec \Phi)^2$ 
that would lift the Goldstones, and preclude 
global monopoles altogether even in the classical limit. It seems interesting 
to explore this further. 

\vskip.75cm

{\bf Acknowledgments}: 
We thank Guido D'Amico, Roberto Emparan, 
David E. Kaplan, Surjeet Rajendran, Alexander Vilenkin, Grigory Volovik and Alexander Westphal for interesting 
conversations and criticism. We are grateful to CERN Department of Theoretical Physics and 
to DESY Theory Group for kind hospitality during the course of
the work reported here. This work is supported in part by the DOE Grant DE-SC0009999.

\end{document}